\documentclass[USenglish,onecolumn]{article}

\usepackage[utf8]{inputenc}
\usepackage[numbers]{natbib}
\usepackage[big]{dgruyter}
\usepackage{color}

\newcommand{\nc}{\newcommand}
\nc{\nn}{\nonumber}

\begin{document}

\articletype{Research Article{\hfill}Open Access}

\author*[1]{Yonatan~Sivan}

\author[2]{Shi-Wei~Chu}

\affil[1]{Unit of Electro-optics Engineering, Faculty of Engineering Sciences, Ben-Gurion University of the Negev, P.O. Box 653, Israel, 8410501, E-mail: sivanyon@bgu.ac.il. }

\affil[2]{Department of Physics, National Taiwan University, No. 1, Sec. 4, Roosevelt Rd., Taipei 10617, Taiwan, E-mail: swchu@phys.ntu.edu.tw. }

\title{\huge Nonlinear plasmonics at high temperatures}

\runningtitle{Nonlinear plasmonics at high temperatures}

\begin{abstract}
  {We solve the Maxwell and heat equations self-consistently for metal nanoparticles under intense continuous wave (CW) illumination. Unlike previous studies, we rely on {\em experimentally}-measured data for the metal permittivity for increasing temperature and for the visible spectral range. We show that the thermal nonlinearity of the metal can lead to substantial deviations from the predictions of the linear model for the temperature and field distribution, and thus, can explain qualitatively the strong nonlinear scattering from such configurations observed experimentally. We also show that the incompleteness of existing data of the temperature dependence of the thermal properties of the system prevents reaching a quantitative agreement between the measured and calculated scattering data. This modelling approach is essential for the identification of the underlying physical mechanism responsible for the thermo-optical nonlinearity of the metal and should be adopted in all applications of high temperature nonlinear plasmonics, especially for refractory metals, both for CW and pulsed illumination.}
\end{abstract}
  \keywords{Thermo-plasmonics, nonlinear optics, metal nanoparticles.}

  \journalname{Nanophotonics}

  \startpage{1}

\maketitle

\section{Introduction}

Nano-plasmonic systems have been intensively studied in recent decades due to their unique potential for local field enhancement and subwavelength confinement, and are considered as promising candidates for a wide variety of applications~\cite{MTM_review_Brongersma_2010,MTM_Review_Barnes}. However, the inherent absorption in the metal proves to be a substantial obstacle towards the realization of real-life applications.

Accordingly, in recent years the applied plasmonic research focused on applications that {\em exploit} the absorption in the metal as means to generate heat on the nanoscale~\cite{Govorov_thermoplasmonics,thermo-plasmonics-review}, a research topic usually referred to as thermo-plasmonics. This resulted in a wide range of emerging applications, at different ranges of temperatures, starting from photothermal (PT) imaging~\cite{PT_imaging,PT_imaging_Zharov}
, through cancer treatment~\cite{PT_treatment}, temperature measurement~\cite{Uriel_T_imaging}, plasmonic photovoltaics~\cite{plasmonic_photovoltaics} and water boiling, sanitation and super-heating~\cite{Baffou-bubble1,Halas-bubble1,Halas-bubble2,Munjiza_2014}, up to thermo-photovoltaics~\cite{Zubin_thermophotovoltaics,thermal_emission_vlad}, diffusive switching~\cite{Khurgin-diffusive-switching}, thermoelectrics~\cite{nano-heat-transfer}, 
plasmon-mediated photocatalysis~\cite{plasmonic_photocatalysis_1,plasmonic_photocatalysis_2,plasmonic_photocatalysis_3}, plasmon-assisted chemical vapor deposition~\cite{plasmonic_cvd} 
and heat-assisted magnetic recording~\cite{refractory_plasmonics}, which may involve temperatures even higher than $2000 ^\circ $K.


In the majority of works in thermo-plasmonics, the optical and thermal properties are assumed to be fixed. However, as the heat is induced by laser illumination (unlike external heating assumed in thermal emission engineering and nanoscale radiative heat transfer~\cite{thermal_emission_Johnson_1,Kapitza-NPs}), it is necessary to account for the coupling between the electromagnetic fields, the temperature, the optical properties and the thermal properties (i.e., heat capacity, thermal conductivity, Kapitza resistance) of the constituents in order to achieve quantitative understanding of the field and temperature distribution. To the best of our knowledge, such a systematic, self-consistent study was not performed so far in the context of thermo-plasmonics. Specifically, the temperature dependence of the metal permittivity was accounted for in some studies, either through a (cubic) thermal nonlinearity~\cite{Hache-cubic-metal-nlty,Boyd-metal-nlty} or more generally, based on a combination of the Two Temperature model~\cite{Two_temp_model} with a complex model of the permittivity. Such models have to correctly account for a rather large number of competing effects within the metal, including electron scattering, thermal expansion, band shifting, the effect of Fermi smearing on intraband transitions~\cite{Hache-cubic-metal-nlty} and on the interband transitions~\cite{Rosei_Au_diel_function,Rosei_Au_diel_function2,Rosei_metal_diel_function}, and more. To the best of our knowledge, such a comprehensive study was done only in~\cite{Stoll_review}. Moreover, during intense illumination, all these effects are modified due to the deviation of the electron distribution function, scattering rates etc. from their equilibrium values, effects which were studied only partially~\cite{Hache-cubic-metal-nlty,non_eq_model_Lagendijk,Seidman-Nitzan-non-thermal-population-model,hot_es_Atwater}. Most importantly, these studies, and the many less complete ones, were all dedicated to the ultrafast regime, such that efforts were made to avoid the longer-term (i.e., few picoseconds and longer) thermal effects (e.g., by looking at nanoparticles not larger than a few nms in diameter, lowering the repetition rate etc., see discussion in~\cite{Hamanaka-Nakamura-2003}) to the extent that the study of these longer-term thermal effects themselves was neglected. Thus, the relative weight of the above-mentioned effects was not studied so far for continuous wave (CW) illumination. However, the study of the CW limit becomes interesting again, with the growing interest in thermo-plasmonic applications.

In the few studies dedicated to longer-term thermal effects (i.e., for CW illumination)~\cite{high-T-plasmonics-Italians-1,high-T-plasmonics-Italians}, the model used for the temperature dependence of the metal permittivity relied on a problematic model that neglected some of the dominant physical effects, most prominently, the temperature dependence of the interband transitions. Even in the mid-infrared (NIR) regime, where no intense sources are available, it did not include all dominant effects known from the literature. Thus, the quantitative predictions in these studies are questionable.

Finally, the temperature dependence of the thermal properties was not taken into account in any of these studies (ultrafast and CW), to the best of our knowledge.

In order to close this knowledge gap, in this article, we perform a thorough theoretical study of the high temperature regime of nano-plasmonic systems under intense optical illumination in the visible range and for CW illumination. Our study focuses on the (classical) interaction between the temperature, permittivity and electromagnetic fields; we use experimentally-measured data for the various optical and thermal properties in order to avoid the need to dwell into the details on the underlying physics, which, as explained, is only partially understood; however, where possible, we try to identify the relevant physical mechanisms by comparing the theory to experimental results. Specifically, we will show that the thermo-optical nonlinearity can be very high, and thus, our study allows us to explain the experimental observations of the strong nonlinear scattering from metal nanoparticles~\cite{plasmonic-SAX-ACS_phot,plasmonic-SAX-PRL,plasmonic-SAX-OE,SUSI}. More generally, this study should serve as the starting point for further experimental and theoretical studies of the underlying physics, of other regime of parameters (specifically, of pulsed illumination, different materials, geometries etc.), and enable a quantitative study of the various applications mentioned above, as well as several others such as nonlinear composites/metamaterials~\cite{Smith_Boyd_EMT_kerr_media,Liao_OL_1998,Hamanaka-Nakamura-2003,Palpant_EMT_kerr_media}, optical limiting~\cite{Elim_APL_2006,Gurudas_JAP_2008,West_JPCB_2003,IOSB}, plasmon lasing~\cite{oulton-nature,Sivan-OE-gain-2009,Xiao-nature} and super-resolution techniques based on metal nanoparticles~\cite{PT_imaging_Zharov,plasmonic-SAX-ACS_phot,plasmonic-SAX-PRL,plasmonic-SAX-OE,SUSI,PT_PA_imaging_Zharov,NP-STED-ACS-NANO,NP-STED-APL,NP-STED-Perspective,NP-STED-Experiment1}. 

The paper is organized as follows. We start by explaining why the temperature dependence of the optical and thermal properties is usually neglected and identify cases where the temperature dependence and mutual coupling of the Maxwell and heat equations is non-negligible. We then solve the Maxwell and heat equations self-consistently for small metal spheres illuminated by intense visible light, and elucidate the large errors in the calculations of the temperature and field distributions associated with neglecting the temperature dependence of the gold permittivity both in- and off-resonance. We then show that the temperature dependence of some additional parameters, such as the thermal conductivity and the Kapitza resistance, is also required for a correct quantitative prediction of the temperature and field distributions. Finally, we discuss the implications of our results to previous experimental work and specify several future measurements necessary for further studies of the strong temperature nonlinearity of metals.

\section{Self-consistent calculation of the temperature in metal nanostructures}\label{sec:self-consist}

We would like to calculate the scattering of an incident continuous wave (CW) off some metal-dielectric nanostructure as a function of pumping intensity and/or temperature. This requires us to understand how much does the metal temperature increase under this illumination and how much, in turn, does this temperature increase affect the metal permittivity, hence, the electromagnetic field distribution around the nanostructure.

In the simplest model, the metal-dielectric system is assigned a single, spatially non-uniform temperature $T$, i.e., we neglect the difference between the electron and lattice temperatures. 
Then, under CW illumination (with no temporal pump modulation), the heat equation governing the temperature dynamics reduces to the Poisson equation,
\begin{equation}\label{eq:Poisson_eq}
\nabla \cdot \left[\kappa(T(\vec{r})) \nabla T(\vec{r})\right] = - p_{abs}(\vec{r};T(\vec{r})),
\end{equation}
where $\kappa$ is the thermal conductivity (specifically, $\kappa_m$ and $\kappa_{host}$ for the metal nanostructure and the dielectric host, respectively). Note that in principle, the thermal conductivity can be temperature-dependent. 

The typical boundary conditions accompanying Eq.~(\ref{eq:Poisson_eq}) are the continuity of the temperature $T$ and heat flux $\kappa \nabla T$ across the interface between the different materials\footnote{However, see~\cite{Baffou_pulsed_heat_eq_with_Kapitza} and the discussion on the Kapitza resistance below. }.

The heat source, $p_{abs}$, represents the density of absorbed power (in units of $W/cm^3$). Classically, the relation between the absorbed energy and local incident electromagnetic field intensity is given by 
\begin{equation}\label{eq:q}
p_{abs}
= \frac{\epsilon_0 \omega}{2} \epsilon''_m
|\vec{E}|^2, 
\end{equation}
where $\vec{E}$ is the total (local) electric field (namely, the solution of the vectorial Helmholtz equation), $\omega$ is the pump frequency and $\epsilon_m = \epsilon_m' + i \epsilon_m''$ is the complex (relative) permittivity of the metal, which serves as the heat source in this problem. This expression is sometimes replaced by $\sim \alpha I$ (or $\sim \sigma_{abs} I_{pump}$)
where $\alpha$ is the absorption coefficient (absorption cross-section) 
and $I$ is the beam intensity. Quantum mechanically, $p_{abs}$ has to be calculated as the spectral integral over all the possible transitions from electron levels and over the frequency content of the pump pulse times the photon energy, involving both absorption and emission to and from each level~\cite{Seidman-Nitzan-non-thermal-population-model}.

The common approach in heat calculations of plasmonic systems (frequently, referred to as thermo-plasmonics~\cite{thermo-plasmonics-review}) is to solve the Maxwell equations first for ambient conditions, i.e., assuming $\epsilon_m(\omega;T = T_{env})$ where $T_{env}$ is the temperature far away from the heat generating (metal) objects. Then, one substitutes the resulting electric field distribution into the heat source term $p_{abs}$~(\ref{eq:q}) in the heat equation~(\ref{eq:Poisson_eq}). Below, we refer to this approach as the 
temperature-independent permittivity (TIP) model.

The TIP approach is appropriate as long as the relative change of of the permittivity, $\Delta \epsilon_m \sim \Delta T\ d\epsilon_m/dT$, is small. Typically, the thermo-derivative, $d\epsilon_m/dT$, varies between $\sim 10^{-5}/^\circ K$ for standard dielectric materials~\cite{Boyd-book} up to $10^{-4}-10^{-3}/^\circ K$ for water~\cite{Shaked-PT_imaging} or metals~\cite{Wilson_deps_dT,Stoll_environment}. Thus, as long as the temperature increase (with respect to room temperature) is modest, i.e., limited to a few degrees, the change of the permittivity is indeed negligible. 
Potentially, opposite signs of the thermo-derivative of the dielectric material and metal may cause the overall temperature dependence of the system under consideration to be weaker than in each of the constituents~\cite{Stoll_environment}, thus, providing further justification for treating the permittivity as temperature-independent. The permittivity changes may be negligible also in the wavelength regime for which $d\epsilon_m/dT$ vanishes. Peculiarly, it turns out that for gold, this regime is around $520-550$nm~\cite{Wilson_deps_dT,Stoll_environment}, i.e., it coincides with the plasmon resonance wavelength of small metal nanospheres, which have been subject to extensive study~\cite{thermo-plasmonics-review,Stoll_review,plasmonic-SAX-ACS_phot,plasmonic-SAX-PRL,plasmonic-SAX-OE,Stoll_environment}. 

However, in plasmonic nanostructures under {\em intense} illumination (as for e.g., thermo-photovoltaics~\cite{Zubin_thermophotovoltaics,thermal_emission_vlad}, plasmon-mediated photocatalysis~\cite{plasmonic_photocatalysis_1,plasmonic_photocatalysis_2,plasmonic_photocatalysis_3}, plasmon-assisted chemical vapor deposition~\cite{plasmonic_cvd} and heat-assisted magnetic recording~\cite{refractory_plasmonics}), the conditions prescribed above are typically {\em not} fulfilled. Indeed, while the (relative) modification of the {\em real} part of the metal permittivity due to changes of the temperature may be small, a substantial increase of the temperature (a few 10's of degrees or more) may cause the {\em imaginary} part of the metal permittivity, $\epsilon_m''$, to change {\em substantially}. 

In the case of {\em external} heating (and weak illumination), one has to use the appropriate permittivity data for the ambient temperature and solve only the Maxwell equations, as done routinely for room temperature studies. In contrast, intense (laser) illumination will result in mutual coupling of the heat and Maxwell equations via $p_{abs}$~(\ref{eq:q}), requiring them to be solved simultaneously. In these cases, the standard model described above (TIP model), which does not take into account the thermo-optical (nonlinear) response to the electromagnetic field, would have to be replaced with a {\em temperature-dependent permittivity} (TDP) model. This is essential in order to make the results from such high applications quantitatively relevant.

Remarkably, it is a common practice to take into account the thermal nonlinearities of the {\em host} medium (e.g., for photothermal imaging~\cite{PT_imaging,PT_imaging_Zharov,PT_PA_imaging_Danielli}, cancer treatment~\cite{PT_treatment}, thermal lensing~\cite{Bonner_thermal_lensing} etc.). However, the majority of studies within the plasmonics community, {\em ignore} the temperature dependence of the metal permittivity. Some of the earlier studies did account for the thermal response of the metal by approximating it with a cubic nonlinearity, see~\cite{Hache-cubic-metal-nlty} or~\cite{Boyd-metal-nlty} for a more recent review. This approach was used, however, only for cases where the pump pulse was not longer than a few nanoseconds, and in the perturbative regime, i.e., where the relative permittivity changes were small such that the cubic approximation is sufficient. Moreover, these studies focused primarily on the electric field distribution, and ignored the temperature itself. Similarly, studies of effective medium theories applied to media with (thermal) cubic nonlinearities also focused on the field rather than the temperature distribution, see e.g.,~\cite{Hache-cubic-metal-nlty,Hamanaka-Nakamura-2003,Smith_Boyd_EMT_kerr_media,Smith_Boyd_EMT_kerr_media_1999,Liao_OL_1998,Elim_APL_2006,Gurudas_JAP_2008,West_JPCB_2003,IOSB,Palpant_EMT_kerr_media} and references therein. One of the reasons for that is obviously that measuring the temperature in the near field of the NPs remains a very difficult task, despite the progress made recently~\cite{Kawata_T_mesurement,NV_centers_T_measurement_NTU}.

Within a TDP model, we expect to be able to distinguish between two scenarios. In the general scenario, as the temperature (and, hence, the imaginary part of the metal permittivity) changes, the heat generation rate $p_{abs}$ changes as well. In particular, if $\epsilon_m''$ grows with temperature, the TIP model will provide an {\em under}-estimate of the actual temperature, as could be calculated from the fully-coupled (TDP) model. The field distribution, on the other hand, may differ only slightly from the field distribution predicted by the simplified TIP model, since typically, $\epsilon_m'' \ll |\epsilon_m'|$.

In contrast, at plasmon resonance, the metal nanostructure acts as a cavity whose quality factor scales inversely with the imaginary part of the metal permittivity, $\epsilon_m''$. Accordingly, if $\epsilon_m''$ increases with temperature, the local electric field (hence, the heat power dissipation) {\em drops}, the resonance broadens, and the overall power dissipation {\em decreases} with respect to the prediction of the TIP model. Accordingly, the temperature will rise more slowly, such that the simplified TIP model will provide an {\em over}-estimate of the actual temperature. In this case, the TDP model will also predict a substantial change of the scattered field with respect to the prediction of the TIP model. The opposite will happen if $\epsilon_m''$ decreases with temperature.

These two effects will be demonstrated analytically and numerically in Section~\ref{sec:spheres}. However, in the meantime, from this discussion, it is obvious that there is strong spectral sensitivity, however, the solution would almost never follow the predictions of the TIP model. Below, we demonstrate the differences between these models in some specific examples, showing that they can be substantial for realistic cases and for many applications which are studied extensively these days.

In order to quantify these differences, one has to have available comprehensive data of the temperature dependence of the metal permittivity. However, quite surprisingly, such data hardly exists, even for gold, which is the plasmonic material studied most extensively, see detailed discussion in~\cite{PT_Shen_ellipsometry_gold}. In the absence of elaborate experimental data, theoretical models for the temperature dynamics~\cite{Two_temp_model,non_eq_model_Lagendijk,non_eq_model_Carpene} and metal permittivity dynamics~\cite{Stoll_review} were developed. However, as mentioned above, effectively all these studies focussed on the {\em ultrafast} (up to a few picoseconds) regime, and only a few of these studies accounted for all the relevant physical mechanisms~\cite{Stoll_review}. Similarly, the multitude of models where the thermal response is approximated as a cubic nonlinearity~\cite{Hache-cubic-metal-nlty,Boyd-metal-nlty} did not consider the thermal response on time scales longer than a few nanoseconds. The quantitative predictions in the few studies of the CW nonlinear response should, as mentioned, be taken with a grain of salt, due to missing ingredients in the permittivity models employed.

Thus, to the best of our knowledge, there is no complete model for the {\em slow} thermal response, as appropriate for CW illumination. In this regime, the electronic response which dominates the ultrafast response becomes negligible, and other effects such as lattice heating and thermal expansion~\cite{Stoll_review}, stress and strain, band shifting~\cite{Ashcroft-Mermin} and indirect (i.e., phonon-assisted) interband transitions~\cite{Pells-Shiga}, take dominance. 

\begin{center}
\begin{figure}[h!]
  \includegraphics[width=11cm]{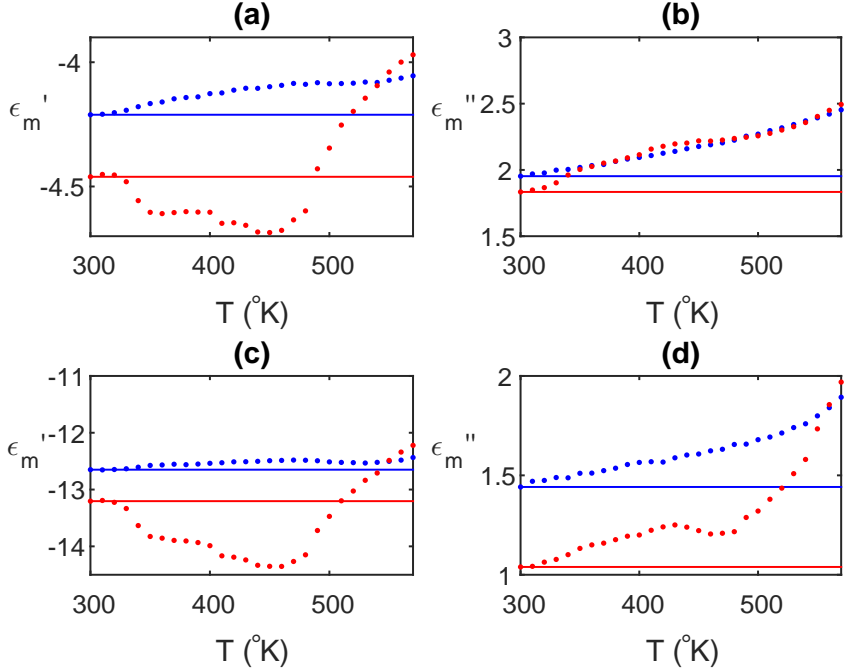}
  \caption{(Color online) (a) Real and (b) imaginary part of the (relative) permittivity extracted from the ellipsometry measurements of an annealed (blue) and unannealed (red) Au film at $\lambda = 533$nm for $300-570 ^\circ$K. (c)-(d) 
  Same data for $\lambda = 671$nm
  . } \label{fig:epsilon}
\end{figure}
\end{center}

In order to close this knowledge gap, we have recently performed ellipsometry measurements to retrieve the permittivity data of bulk gold at increasing temperatures~\cite{PT_Shen_ellipsometry_gold}. Our study showed that $\epsilon_m''$ increases substantially with temperature across the visible spectral range; Indeed, in the temperature regime of $300-570 ^\circ$K, Fig.~\ref{fig:epsilon} shows changes of up to $\sim 25-30\%$ in the visible range for selected wavelengths. In the near IR regime, an increasingly stronger dependence on temperature was observed~\cite{PT_Shen_ellipsometry_gold}.

Our study also showed that the changes to the real part of the gold permittivity are substantially smaller with respect to the room temperature values - $\sim 1-3\%$ in the temperature regime studied here. Similar results appear in two recent independent studies~\cite{Taiwanese_ellipsometry_2014,Shalaev_ellipsometry_gold} as well as for Ag~\cite{Taiwanese_ellipsometry_2014,Indian_Ag_ellipsometry_2014}; some of these papers also offer a fit of the retrieved data to a Drude-Lorentz model.

Thus, in order to simplify the modelling and discussion, we assume in what follows that the host (dielectric) material is purely real and non-dispersive. This assumption has a negligible effect on our results. Indeed, the numerical examples shown below show that the changes of the host permittivity and of the real part of the metal permittivity have a secondary effect on the temperature and field distribution. This residual temperature dependence will have to be taken into account in applications of photothermal imaging~\cite{PT_imaging,PT_imaging_Zharov,PT_PA_imaging_Danielli} and treatment~\cite{PT_treatment}, water boiling~\cite{Halas-bubble1,Halas-bubble2}, plasmonic (thermo-)photovoltaics~\cite{Zubin_thermophotovoltaics,thermal_emission_vlad} and thermal lensing~\cite{Bonner_thermal_lensing}, plasmon-assisted catalysis~\cite{plasmonic_photocatalysis_1,plasmonic_photocatalysis_2,plasmonic_photocatalysis_3}, where the level of temperature rise of the surrounding medium is critical. This is, however, left for future studies. 



\section{Metal spheres}\label{sec:spheres}

Although in general, the problem at hand requires a self-consistent numerical iteration scheme involving both the Maxwell and heat equations~\cite{high-T-plasmonics-Italians-1,high-T-plasmonics-Italians}, for some simple geometries, one can avoid solving the Maxwell equations and rely on known solutions. 
As a generic example, we now consider the temperature of a single small ($a \ll \lambda$) spherical metal nanoparticle (NP) illuminated by a plane wave. Hence, the quasi-electrostatic solution of Maxwell equations~\cite{Bohren-Huffman-book} holds such the electric field inside the nanoparticle ($\vec{E}_{NP} \equiv \vec{E}(r < a)$) is uniform and given by
\begin{eqnarray}
\vec{E}_{NP} = \frac{3\epsilon_d}{\epsilon_m(\omega;T) + 2 \epsilon_d} \vec{E}_{inc}, \label{eq:E_NP_q-static}
\end{eqnarray}
where $\vec{E}_{inc}$ is uniform, i.e., a fixed parameter; in what follows, we suppress the vector symbol. Using Eq.~(\ref{eq:E_NP_q-static}) in Eq.~(\ref{eq:q}) gives a uniform power dissipation
\begin{equation}\label{eq:q-sphere}
p_{abs}(\omega,T(r<a)) = \frac{\epsilon_0 \omega}{2} \epsilon''_m(T) \left|\frac{3\epsilon_d}{\epsilon_m(T) + 2 \epsilon_d}\right|^2 |E_{inc}|^2. 
\end{equation}
The solution of the Poisson equation~(\ref{eq:Poisson_eq}) for this case is~\cite{thermo-plasmonics-review}
\begin{equation} \label{eq:T_NP} T(r) = T_{env} + \left\{
\begin{array}{cc}
\frac{P_{abs}(\omega,T)}{4 \pi \kappa_{host} a} \left[1 + \frac{\kappa_{host}}{2 \kappa_m}\left(1 - \frac{r^2}{a^2}\right)\right], & \quad \quad r < a, \\
\frac{P_{abs}(\omega,T)}{4 \pi \kappa_{host} r}, & \quad \quad r > a,
\end{array} \right.
\end{equation}
where $\kappa_m$ and $\kappa_{host}$, the thermal conductivities of the NP and host, respectively, are assumed for the moment to be temperature independent (hence, uniform) and $P_{abs}(\omega,T) \equiv \int p_{abs} dV$ is the total power dissipated in the NP.


Since typically $\kappa_m \gg \kappa_{host}$, it follows from Eq.~(\ref{eq:T_NP}) that diffusion is sufficiently strong to homogenize the temperature within the NP; this assumption is supported by exact simulations, showing temperature uniformity even for much larger nanoparticles~\cite{high-T-plasmonics-Italians}. Thus, neglecting the small temperature variation, we can define $T_{NP} \equiv T(r <a)$ so that $P_{abs} = 4 \pi a^3 p_{abs}/3$ and by  Eq.~(\ref{eq:T_NP}), we get
\begin{equation}\label{eq:TTT}
T_{NP} 
= T_{env} + \frac{a^2}{3\kappa_{host}} p_{abs}(T_{NP}).
\end{equation}
Substituting Eq.~(\ref{eq:q-sphere}) in Eq.~(\ref{eq:TTT}) gives
\begin{eqnarray}\label{eq:T_self-consistent}
T_{NP} 
&=& T_{env} + \frac{\epsilon_0 \omega}{2} \frac{3 a^2}{\kappa_{host}} |E_{inc}|^2 \frac{\epsilon_d^2}{\left|\epsilon'_{tot} + \epsilon''_m(T_{NP})\right|^2}  \epsilon''_m(T_{NP}),
\end{eqnarray}
where $\epsilon'_{tot} \equiv \epsilon_m' + 2 \epsilon_d$. 

Eq.~(\ref{eq:T_self-consistent}) is a simple root equation for the NP temperature which is easy to solve. However, before presenting detailed numerical examples, we discuss several general properties of the solution.

In the general (off-resonance) case, the real parts of the permittivities do not perfectly cancel, such that typically, $\epsilon'_{tot} \gg \epsilon_m''$. Then, we get
\begin{eqnarray}\label{eq:T_self-consistent_off}
T_{NP} &\approx& T_{env} + \frac{\epsilon_0 \omega}{2} \frac{3 a^2}{\kappa_{host}} |E_{inc}|^2 \frac{\epsilon_d^2}{\left|\epsilon'_{tot}\right|^2} \epsilon''_m(T_{NP}).
\end{eqnarray}
Indeed, as predicted, we see from Eq.~(\ref{eq:T_self-consistent_off}) that the absorbed power (hence, the overall temperature) will be {\em higher} in the TDP model compared with the TIP model.

On the other hand, at resonance, the real part of the denominator vanishes such that the power dissipation~(\ref{eq:q-sphere}) drops for increasing $\epsilon''_m$. This is due to the resonant nature of the interaction - $\epsilon_m''$ is inversely proportional to the quality factor of this ``effective'' resonator. At the same time, the resonance broadens due to the temperature rise and the temperature itself is given by 
\begin{equation}\label{eq:T_self-consistent_on}
T_{NP} = T_{env} + \frac{\epsilon_0 \omega}{2} \frac{3 a^2}{\kappa_{host}} |E_{inc}|^2 \frac{\epsilon_d^2}{\epsilon''_m(T_{NP})}.
\end{equation}
We thus see from Eq.~(\ref{eq:T_self-consistent_on}) that the absorbed power (hence, the overall temperature) will be {\em lower} in the TDP model compared with the TIP model.

From the above discussion, the reasons for neglecting the changes of the real part of the metal (and dielectric) permittivities become apparent. Indeed, the changes of $\epsilon'_{tot}$ are relatively small, and cause only a slight shift of the plasmon resonance position. 
It is clear, however, that the important parameter is the shift of the real part of $\epsilon'_{tot}$ rather than the shift of any of its constituents, as various combinations of thermo-derivatives of the metal and dielectric will give rise to shifts in different directions.

Once the NP temperature is determined, one can calculate the scattered field using the quasi-static solution~\cite{Bohren-Huffman-book}. 
In the case of a single intense (pump) beam, the scattered field, $\vec{E}_{sc} \equiv \vec{E}(r > a) - \vec{E}_{inc}$, is given by
\begin{eqnarray}
\vec{E}_{sc}(\omega_{pump},T_{NP}) &=& \frac{\epsilon_d - \epsilon_m(\omega_{pump},T_{NP})}{\epsilon_{tot}' + i \epsilon_m''(\omega_{pump},T_{NP})} \frac{a^3}{r^3} \left|E_{inc}(\omega_{pump})\right| \left(2 \cos \theta \hat{r} + \sin \theta \hat{\theta}\right). \label{eq:E_sc_q-static_pump}
\end{eqnarray}
When the intense beam is accompanied by a second, weaker (probe) beam, the scattering of the probe will be given by the same expression, where the permittivities and fields are evaluated at the probe frequency. Since this case is effectively similar to the standard linear case or to the case of external heating~\cite{epsilon-T-dependence-Ukrains,PT_Shen_ellipsometry_gold}, it will not be considered further.

\subsection{Numerical examples}\label{sub:numerics}

\begin{center}
\begin{figure}[h!]
  \includegraphics[width=15cm]{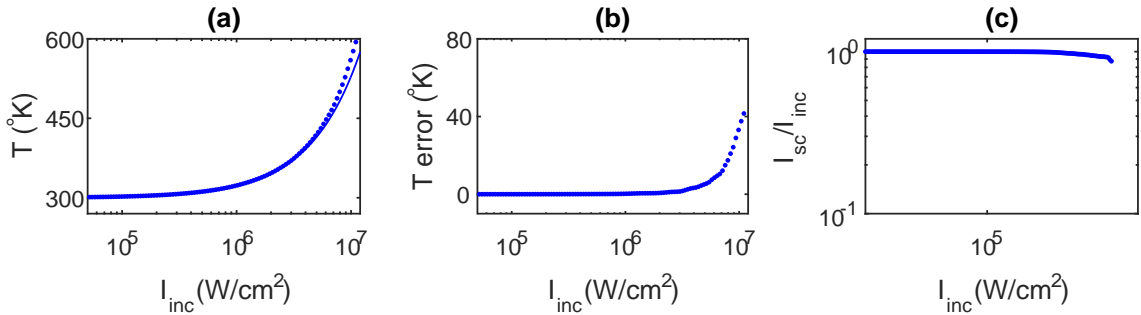}
  \caption{(Color online) (a) Calculated temperature for $\lambda = 533$nm and the off-resonance case ($\epsilon_d = 5.5$) for the TDP model (blue dots) and TIP model (solid blue line) based on annealed permittivity data. (b) The temperature difference between both models. (c) Peak intensity of the scattered field as a function of the incoming pump intensity for the two models. } \label{fig:off-resonance-533}
\end{figure}
\end{center}

Based on the experimental data for annealed gold~\cite{PT_Shen_ellipsometry_gold}, as appropriate for metal NPs made by pulsed laser ablation of gold films~\cite{PLAL}, we initially solve Eq.~(\ref{eq:T_self-consistent}) for $\lambda = 533$nm (permittivity data given in Fig.~\ref{fig:epsilon}(a)-(b)).
Fig.~\ref{fig:off-resonance-533}(a)-(b) show that when the system is tuned away from resonance (host permittivity is $\epsilon_{host} = 5.5$), for sufficiently large pumping intensity, the naive (TIP) model indeed {\em under-estimates} the temperature rise in the particle. For example, we see that the TDP model predicts $T = 594 ^\circ K$, while the TIP gives $T = 552 ^\circ K$, i.e., an error of $\sim 17\%$ of the temperature increase. This error is commensurate with the corresponding change of $\epsilon_m''$ which increased by $\sim 30\%$ (the real part changes by $\sim 1\%$). Nevertheless, the TIP model still predicts the scattered field with reasonable accuracy (Fig.~\ref{fig:off-resonance-533}(c)). Indeed, in this case, the denominator for the expression for the scattering (see Eq.~(\ref{eq:E_sc_q-static_pump})) is only slightly affected by the change of the imaginary part of the permittivity.

\begin{center}
\begin{figure}[h!]
  \includegraphics[width=15cm]{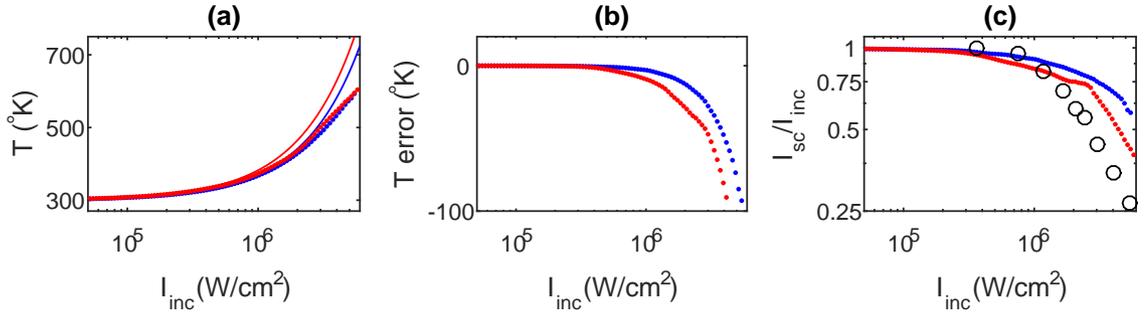}
  \caption{(Color online) Same as in Fig.~\ref{fig:off-resonance-533} for the on-resonance case ($\lambda = 533$nm and $\epsilon_d = 2.25$). Also shown are the results for un-annealed data (red dots) and the (normalized) experimentally-measured data of scattering from a single $40$nm Au NP embedded in index matching oil under CW illumination. } \label{fig:on-resonance-533}
\end{figure}
\end{center}

At resonance, on the other hand ($\lambda = 533$nm, but with $\epsilon_d = 2.25$), the naive (TIP) model {\em over-estimates} the temperature rise in the NP (see Fig.~\ref{fig:on-resonance-533}(a)-(b)). For example, when the TDP model predicts $T = 594^\circ K$, the TIP gives $T = 694 ^\circ K$, i.e., $\sim 34\%$ error in temperature rise measurement. More importantly, in this case, the TDP model predicts a $40\%$ decrease of the scattering (Fig.~\ref{fig:on-resonance-533}(c)). Fig.~\ref{fig:on-resonance-533} also shows the results based on un-annealed permittivity data, as appropriate for metal NPs synthesized in solution~\cite{PT_Shen_ellipsometry_gold}. One can see that while the results are qualitatively similar, the non-annealed gold shows a much stronger sensitivity to the rising temperature. This emphasizes the need to account for the relevant permittivity data depending on the metal particle preparation method~\cite{PT_Shen_ellipsometry_gold}.

Most importantly, Fig.~\ref{fig:on-resonance-533}(c) also shows a comparison to measured scattering data\footnote{See discussion of the experimental setup in~\cite{plasmonic-SAX-PRL}. } from a single $40$nm Au NP embedded in index matching oil under CW illumination. One clearly observes very good qualitative agreement between the theory and the measurement, achieved without any fitting parameters. This agreement between the scattered fields also reveals the NP temperature - the lowest scattering levels are attained for a temperature rise of only a few hundreds of degrees, i.e., well below the melting temperature of the Au NPs, which is somewhat less than $1000 ^\circ$K~\cite{Munjiza_2014}. Note that a quantitative agreement of the scattered field and temperature requires further refinement of our modelling, see discussion Section~\ref{sec:discussion}.

\begin{center}
\begin{figure}[h!]
  \includegraphics[width=15cm]{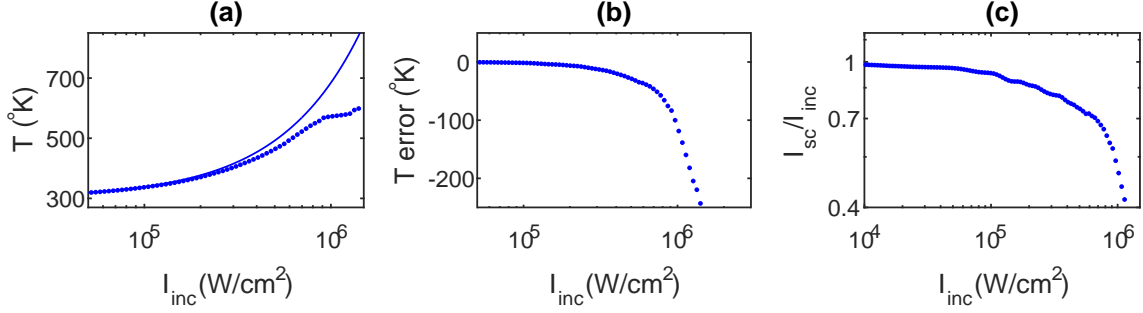}
  \caption{(Color online) Same as Same as in Fig.~\ref{fig:off-resonance-533} for the on-resonance case with $\lambda = 671$nm and $\epsilon_d = 2.5$. } \label{fig:on-resonance-671}
\end{figure}
\end{center}

As a comparison, we show in Fig.~\ref{fig:on-resonance-671} the temperature and scattered field for resonant illumination at $\lambda = 671$nm ($\epsilon_d = 2.5$); permittivity data at this wavelength is shown in Fig.~\ref{fig:epsilon}(c)-(d). While the trends are qualitatively similar to the case of $\lambda = 533$nm, the nonlinear response is stronger - the scattered field drops by $60\%$ and the temperature error is up to $\sim 250 ^\circ$K.

Finally, we note that when the variation of the real part with temperature is taken into account (i.e., when we solve Eqs.~(\ref{eq:T_self-consistent}) and~(\ref{eq:E_sc_q-static_pump}) for $\epsilon_{tot}'(T) \ne 0$), there is no substantial change of any of the results described so far. This happens because the (absolute, as well as) relative changes of the real part of the metal permittivity with temperature are typically smaller than those of the imaginary part; accordingly, they have far smaller influence on the temperature and scattering.

\section{Additional considerations}\label{sec:additional}
Below, we discuss two additional aspects of the therm-optical problem at hand that, to the best of our knowledge, are discussed for the first time in the current context.

\subsection{The role of the temperature dependence of the thermal conductivity}\label{sub:kappa}

So far, we assumed that the thermal conductivities are temperature-independent. However, the temperature-dependence of the thermal conductivity is well-known for a wide range of materials. Remarkably, its variation with temperature is comparable to that of the metal permittivity. For example, the thermal conductivity of water increases by about 10\% between $300$ to $400 ^\circ$K; beyond this temperature, the water boils. Oil exhibits comparable changes over a wider temperature range, with some oils exhibiting increased conductivity with growing temperature, and some exhibiting reduced conductivity. The thermal conductivity of other materials, like collagen~\cite{kappa_T_biology}, quartz
, silicon wires or aluminum oxide 
exhibit even stronger temperature dependence. The thermal conductivity of the metal itself also varies substantially with the temperature, however, since it is typically much larger than the host conductivity, this variation plays a negligible role for our purposes (see Eq.~(\ref{eq:T_NP})). Thus, it is clear that this dependence has to be taken into account in order to accurately determine the temperature and field distributions. In general, if the host thermal conductivity increases with temperature, then the temperature rise is lower than that predicted by a model that ignores this effect, and vice versa.

\begin{center}
\begin{figure}[h!]
  \includegraphics[width=6cm]{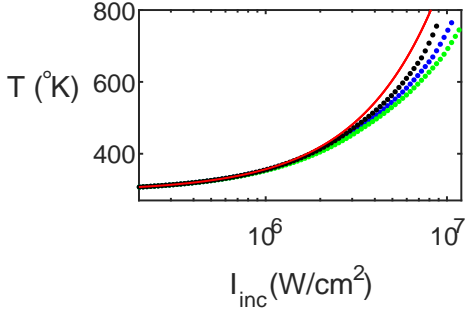}
  \caption{(Color online) Calculated temperature for on-resonance case ($\lambda = 533$nm) for the TIP model (solid red line) and TDP model with temperature-independent thermal conductivity (blue dots), both as in Fig.~\ref{fig:on-resonance-533}, compared with a thermal conductivity that increases with temperature (green dots) and thermal conductivity that decreases with temperature (black dots). In the latter two cases, the thermal conductivity changes by as much as $13\%$. } \label{fig:kappa}
\end{figure}
\end{center}

The exact solution~(\ref{eq:T_self-consistent}) used so far will not hold anymore for a temperature- (hence, space-)~dependent thermal conductivity. However, exploiting again the uniformity of the temperature inside the NP, allows us to keep using the implicit relation~(\ref{eq:T_self-consistent}). Numerical simulations, see Fig.~\ref{fig:kappa}, show that the error associated with the change of the thermal conductivity with the temperature (taken as $T_{env}/\kappa_{env} d\kappa/dT \sim \pm 10\%$) is of the same order of the temperature change itself. As expected, a similar trend is found also for the off-resonant case (not shown).

\subsection{The role of the interface (Kapitza) conductivity} \label{sub:Kapitza}

A more realistic model of the heat transfer between the NP and its surrounding has to account for the finite interface (Kapitza) conductivity, $g$~\cite{Munjiza_2014,Kapitza-NPs,Kapitza-review}. 
In this case, it was shown~\cite{Baffou_pulsed_heat_eq_with_Kapitza} that the solution for the illuminated sphere is modified only inside the sphere, namely,
\begin{equation} \label{eq:T_NP_Kapitza}
T_{NP} = \frac{P_{abs}(\omega)}{4 \pi a} \left[\frac{1}{\kappa_{host}} + \frac{1}{g a}\right],
\end{equation}
where, for simplicity, we again assumed a uniform temperature inside the NP. The finiteness of the interface Kapitza conductivity means that the generated power within the NP escapes more slowly, hence, the overall NP temperature is higher (with respect to the case of infinite interface conductivity).

In general, the value of the interface Kapitza conductivity is known only for a select few cases - its calculation requires heavy and somewhat ambiguous molecular dynamics simulations (see e.g.,~\cite{Munjiza_2014} for a discussion) and its measurement is a tough task. However, fortunately, it turns out that gold nanostructures were some of the few cases that were studied. A fit to experimental results performed in~\cite{Stoll_environment} yielded $g \sim 110 MW/m^2 K$ for the interface between a $18$nm gold sphere and water. With this value, the correction in Eq.~(\ref{eq:T_NP_Kapitza}) with respect to the case of infinite interface conductivity is $\kappa_{host}/g a 
= 33nm/a$. Thus, for the small particles under consideration here, this term is clearly far from being negligible. A similar procedure for gold-ethanol interface yielded $g \sim 40 MW/m^2 K$, i.e., again, providing a substantial contribution. Similar values were reported for the Kapitza conductance between gold and silicon under various surface treatments at temperatures below room temperature~\cite{Kapitza-Au-silicon}.

If the Kapitza conductivity was studied in only a limited number of papers, then its temperature dependence was studied even less. Molecular dynamics calculations performed in~\cite{Munjiza_2014} for $3$nm gold NPs yielded $g \sim 180 MW/m^2 K$ and a temperature dependence similar to that of the permittivity and the thermal conductivity, i.e., a variation by more than $10\%$ for the temperature range covered in the current manuscript ($300-800 ^\circ K$). This temperature dependence has a similar effect to that of the thermal conductivity - an increasing conductance with temperature will give rise to lower temperatures compared with models that ignore it.

\section{Discussion}\label{sec:discussion}
The results shown above raise a clear need to take into account the temperature dependence of the optical and thermal properties of the metal (and its surroundings) in calculations of field and temperature under intense illumination conditions. In particular, the errors associated with the neglect of the temperature dependence of these quantities grow monotonically with the temperature rise, and can reach several 10's or even 100's of degrees for the refractory applications, i.e., even up to $100\%$ relative errors; for resonant illumination, there are comparable relative errors in the scattered fields. In fact, for some applications, such as PT imaging~\cite{PT_imaging,PT_imaging_Zharov,PT_PA_imaging_Danielli}, correcting errors of even a few percent could be substantial.

More generally, our calculations provide a complete treatment of nonlinear plasmonic systems at the high temperature regime that goes beyond the perturbative description of the thermo-optical response~\cite{Hache-cubic-metal-nlty,Boyd-metal-nlty,Wilson_deps_dT}. Indeed, we intentionally avoid any assumption on the functional dependence of the metal permittivity on the temperature or intensity (e.g., an assumption of a cubic nonlinear response~\cite{Hache-cubic-metal-nlty}, or of a constant thermo-derivative $d\epsilon_m/dT$~\cite{Palpant_EMT_kerr_media,Wilson_deps_dT}, or of an averaged response, in the effective medium spirit~\cite{Palpant_EMT_kerr_media}). This approach allowed us to identify the thermo-optical mechanism as being responsible for the nonlinear scattering of monochromatic waves from Au NPs that was observed experimentally~\cite{plasmonic-SAX-ACS_phot,plasmonic-SAX-PRL,plasmonic-SAX-OE,SUSI}, showing deviations from the linear prediction (TIP) of several tens of percent, see Fig.~\ref{fig:on-resonance-533}. Indeed, such changes of scattering are shown to be commensurate with the change of the imaginary part of the metal permittivity with the temperature, see Fig.~\ref{fig:epsilon}. Remarkably, we confirm that the effect occurs on a subwavelength scale - from a {\em single} nanoparticle and potentially its immediate surrounding (via the thermal conductivity), rather than being an effect accumulated on macroscopic distances or due to inter-particle interactions or aggregation, as one may conclude from previous studies of nanoparticle suspensions, see e.g.,~\cite{Liao_OL_1998,Elim_APL_2006,Opt_Limiting_NUS_2008,West_JPCB_2003,IOSB,Gurudas_JAP_2008}. 

In contrast to previous works, which relied on a theoretical model that missed some dominant physical effects~\cite{high-T-plasmonics-Italians-1,high-T-plasmonics-Italians,epsilon-T-dependence-Ukrains}, our study relies on {\em experimentally}-measured permittivity data~\cite{PT_Shen_ellipsometry_gold} and focuses on the visible range. Furthermore, we show stronger effects from NPs smaller than those studied before. Yet, it is important to note that our model provide only a {\em qualitative} match to the experimental data. A quantitative agreement requires accounting for the actual size of the particles (i.e., to go beyond the quasi-static approximation employed here, as done in~\cite{high-T-plasmonics-Italians}) and for the temperature dependence of the thermal properties of the metal and host which is currently not known. For completeness, it is also desired to develop a theoretical model for the (slow) thermal nonlinearity of gold, to support the experimental results. In contrast to the common models (used in some previous studies~\cite{high-T-plasmonics-Italians-1,high-T-plasmonics-Italians,epsilon-T-dependence-Ukrains}), a complete theoretical model will have to account for the temperature dependence of the metal permittivity on both intraband {\em and interband} transitions, and specifically, for the effects of the temperature on the NP volume, electron scattering rates, electron distribution (Fermi smearing), lattice spacing (band shifting), stress/strain build up, as well as for non-equilibrium effects and multi-photon absorption, which we neglected. A model that describes the interplay and relative importance of these effects is yet to be developed. Such a model will be also particularly important in order to explain the nonlinear scattering under {\em pulsed} illumination, which typically involves higher intensities that for CW (up to $GW/cm^2$), and exhibited {\em opposite} trends to those observed for CW illumination~\cite{West_JPCB_2003,Elim_APL_2006,Gurudas_JAP_2008,Opt_Limiting_NUS_2008,IOSB}.

In the same vein, we should mention that the current analysis of the thermal effects may not be sufficient to address the complete intensity-dependence of the scattering from metal nanoparticles. Indeed, it was shown~\cite{plasmonic-SAX-ACS_phot,plasmonic-SAX-OE} that for sufficiently high excitation intensities, the decrease of the scattering changes to a sharp increase, occasionally, and somewhat confusingly, referred to as ``reverse saturation''\footnote{This nomenclature was adopted due to the reminiscence of the absorption/scattering data to that obtained from some atoms or molecules. }. This effect may be related to electron population redistribution due to Fermi smearing (i.e., based on the distribution of thermalized electrons), which shows a rather complicated and non-intuitive spectral dependence with several spectral regimes where the permittivity decreases upon heating~\cite{Rosei_Au_diel_function,Rosei_Au_diel_function2,Rosei_metal_diel_function,Stoll_review}. Alternatively, the increased scattering may be related to absorption saturation (i.e., based on the distribution of {\em non}-thermal electrons)~\cite{Hache-cubic-metal-nlty,Seidman-Nitzan-non-thermal-population-model}, or to an effect associated with the host (e.g., (nonlinear) absorption, phase/structural change etc.). The determination of its origin also awaits the comprehensive permittivity model, and thus, left for a future study.

In that regard, we emphasize that the use of the permittivity data under external heating in laser-illumination calculations (as adopted in the current study or in~\cite{Stoll_review}) is justified only if effects associated with non-thermal electrons, which accompany intense illumination, are negligible compared with effects associated with thermalized electrons. This seems to be the case for gold~\cite{Boyd-metal-nlty} - the absorption saturation due to interband transitions, which is related with partial population inversion, is predicted~\cite{Hache-cubic-metal-nlty} and experimentally verified~\cite{Liao_OL_1998,Smith_Boyd_EMT_kerr_media_1999} to be smaller than the nonlinearity associated with heating (thermalized electrons, Fermi smearing), at least for moderately high intensities and pulsed illumination. A simple estimate based on the measured cubic nonlinearity, ${\chi^{(3)}_{Au}}'' \sim 5 \cdot 10^{-9} cm^2/W$ at $\lambda = 532$nm~\cite{Hache-cubic-metal-nlty,Liao_OL_1998,Smith_Boyd_EMT_kerr_media_1999}, together with the associated field enhancement within the gold, $\left|3 \epsilon_d/\epsilon_m''\right|^2 \sim 20$ (see Eq.~(\ref{eq:E_NP_q-static}) and~\cite{NFE-thermo-plas-FOM}), shows that the permittivity change for $I_{inc} = 1 MW/cm^2$ induces a $\sim 10\%$ change of the imaginary part of the permittivity, as indeed we observe experimentally, see Figs.~\ref{fig:epsilon} and~\ref{fig:on-resonance-533}.\footnote{This implies, specifically, that the arguments used in~\cite{plasmonic-SAX-PRL} to explain the scaling of the deviation point of the scattering curve from the linear prediction were {\em incorrect}. This error originated from the use of the macroscopic absorption cross-section (of the NP) rather than the atomistic cross-section in the estimate of the relaxation time, as appropriate for a single atom effect. Indeed, as the relaxation time in metals is several orders of magnitude faster than in semiconductors, one can expect the saturation levels to be correspondingly higher. Yet, the scaling of the onset of the nonlinear effect is surely a valid experimental observation and is explained in the current context by the thermal effect, which indeed scales with the absorption cross-section of the NP itself, see discussion under Eq.~(\ref{eq:q}). }

Having said that, we emphasize that the estimates above, which are based on measurements of the {\em ultrafast} thermal response, are only partially appropriate for the current context of a CW illumination. Indeed, the ultrafast thermal nonlinearity was derived in~\cite{Hache-cubic-metal-nlty} by neglecting the diffusion of heat from the NP to its surroundings (see also~\cite{Boyd-book}). In our configuration, however, heat diffusion is clearly important (see e.g., Eqs.~(\ref{eq:T_self-consistent})) so that the overall thermal response depends also on the host properties, as well as on the particle size etc.. This may give rise to different values of nonlinearity. In general, though, as already noted above, a complete quantitative match of the model to the experimental data will have to be deferred to a future study.


Finally, we hope that our study would motivate further studies of thermo-optical nonlinearities at the high temperature regime of other gold NPs, as well as similar studies of other metals, especially those proposed for use in refractory plasmonics applications~\cite{refractory_plasmonics}.






{\acknowledgement We would like to thank P.-T. Shen, I. Gurwich, M. Spector and Y. Dubi for many useful discussions. 
Y. Sivan acknowledges financial support from the People Programme (Marie Curie Actions) of the European Union's Seventh Framework Programme (FP7/2007-2013) under REA grant agreement no. 333790. S.W. Chu acknowledges financial support from the Ministry of Science and Technology, Taiwan, under grants number MOST 101-2923-M-002-001-MY3, and MOST-102-2112-M-002-018-MY3.}

%


\bibliographystyle{unsrt}
\bibliography{my_bib}

\end{document}